\documentclass[sigconf]{acmart}

\AtBeginDocument{%
  }
    
\usepackage{balance}
\usepackage{multirow}
\usepackage{stfloats}
\usepackage{booktabs}
\usepackage{array} 

\acmConference[MM'23]{ACM Multimedia}{October29-November 2}{Ottawa, Canada}

\copyrightyear{2023}
\acmYear{2023}
\setcopyright{acmlicensed}\acmConference[MM '23]{Proceedings of the 31st
ACM International Conference on Multimedia}{October 29-November 3,
2023}{Ottawa, ON, Canada}
\acmBooktitle{Proceedings of the 31st ACM International Conference on
Multimedia (MM '23), October 29-November 3, 2023, Ottawa, ON, Canada}
\acmPrice{15.00}
\acmDOI{10.1145/3581783.3612859}
\acmISBN{979-8-4007-0108-5/23/10}

\begin{document}

\title{Hierarchical Audio-Visual Information Fusion with  Multi-label Joint Decoding for MER 2023}

\author{Haotian Wang}
\author{Yuxuan Xi}
\author{Hang Chen}
\author{Jun Du}
\authornote{corresponding author}
\affiliation{%
  \institution{University of Science and Technology of China}
  \city{Hefei}
  \state{Anhui}
  \country{China}
}

\author{Yan Song}
\author{Qing Wang}
\author{Hengshun Zhou}
\author{Chenxi Wang}
\affiliation{%
  \institution{University of Science and Technology of China}
  \city{Hefei}
  \state{Anhui}
  \country{China}
}

\author{Jiefeng Ma}
\author{Pengfei Hu}
\author{Ya Jiang}
\author{Shi Cheng}
\affiliation{%
  \institution{University of Science and Technology of China}
  \city{Hefei}
  \state{Anhui}
  \country{China}
}

\author{Jie Zhang}
\affiliation{%
  \institution{University of Science and Technology of China}
  \city{Hefei}
  \state{Anhui}
  \country{China}
}

\author{Yuzhe Weng}
\affiliation{%
  \institution{Northwestern Polytechnical University}
  \city{Xi'an}
  \state{Shaanxi}
  \country{China}
}

\renewcommand{\shortauthors}{Haotian Wang et al.}

\begin{abstract}
  In this paper, we propose a novel framework for recognizing both discrete and dimensional emotions. In our framework, deep features extracted from foundation models are used as robust acoustic and visual representations of raw video. Three different structures based on attention-guided feature gathering (AFG) are designed for deep feature fusion. Then, we introduce a joint decoding structure for emotion classification and valence regression in the decoding stage. A multi-task loss based on uncertainty is also designed to optimize the whole process. Finally, by combining three different structures on the posterior probability level, we obtain the final predictions of discrete and dimensional emotions. When tested on the dataset of multimodal emotion recognition challenge (MER 2023), the proposed framework yields consistent improvements in both emotion classification and valence regression. Our final system achieves state-of-the-art performance and ranks third on the leaderboard on MER-MULTI sub-challenge. 
\end{abstract}



\maketitle

\section{Introduction}

Multimodal emotion recognition (MER) plays a crucial role in natural human-machine interaction~\cite{HMI,2019er}, intelligent education tutoring~\cite{EduTu2,EduTu4}, and mental health diagnoses~\cite{health1,health2}, etc. In our daily life, when we engage in dialogue and communication, we usually convey our emotions through both verbal and non-verbal content, such as facial expressions and body language~\cite{facial1}. Previous studies focused on emotion recognition in text~\cite{text1,text2}, facial expression~\cite{facial1,facial2} and audio~\cite{audio1,audio2}. However, it has been observed that research on single-modality approaches has reached a certain bottleneck, 
thereby leading to increased attention toward the use of multimodal approaches~\cite{multimodal1,multimodal3,multimodal4,multimodal5}.

Regarding human communication scenarios, emotions are mainly expressed through speech and facial expressions, each providing complementary information. As a result, researchers are dedicated to fusing audio and video modal features~\cite{fuse1, fuse2, fuse3, fuse4}. For example, Han \textit{et al.}~\cite{fuse1} proposed a hierarchical approach that maximized the Mutual Information (MI) among unimodal inputs. Hazarika \textit{et al.}~\cite{fuse2} projected each modality to modality-invariant and modality-specific spaces to learn effective representations. Recently, inspired by the success of pre-trained deep features like wav2vec2.0~\cite{w2v2} and HUBURT~\cite{hubert} in other speech-related tasks, some researchers investigate their superiority over hand-engineered features in speech emotion recognition and discovered that these deep features capture more robust acoustic representations~\cite{wavlm, unimse, SERssl,multilabel2}. 

According to the theory of psychological research, there are two main emotional calculation models: discrete theory and dimensional theory. Discrete theory~\cite{survey2} describes emotional states as discrete labels such as "sad" and "happy". On the other hand, the theory of dimensionality~\cite{dimension} suggests that emotional states exist as points in a continuous space. This allows for simulating subtle, complex, and sustained emotional behaviors. Previous studies have found that there exists a strong correlation between these two models~\cite{multilabel1,multilabel2}. 

In this paper, we propose an efficient multimodal emotion recognition system to recognize both discrete emotion (emotion) and dimensional emotion (valence). Firstly, we extract deep features through different layers of various pre-trained models as robust acoustic and visual representations of raw video segments. Then we fuse these features through three proposed feature fusion structures based on AFG~\cite{AFG}. Afterwards, we have designed a joint decoding module that considers both discrete and dimensional theories based on the correlation between emotion and valence to generate decisions for these two dimensions. A multi-task loss function based on uncertainty~\cite{uncertainty} is also designed to optimize the whole encoding and decoding process. Finally, we generate final predictions of both discrete and dimensional emotions by decision-level fusion. We evaluate the proposed framework on the dataset of the multimodal emotion recognition challenge (MER 2023)~\cite{MER2023}. Experiments show that our framework yields consistent improvements on both emotion classification and valence regression and ranks third on the leaderboard on MER-MULTI sub-challenge.

\section{Methods}

In this section, we will discuss our proposed multimodal emotion recognition system in two subsections. The feature extraction and fusion strategies will be illustrated in the first subsection. The proposed joint decoding module of discrete and dimensional emotions and the designed multi-task loss function will be introduced in the second subsection.
\begin{figure}[t]
  \centering
  \includegraphics[width=1.0\linewidth]{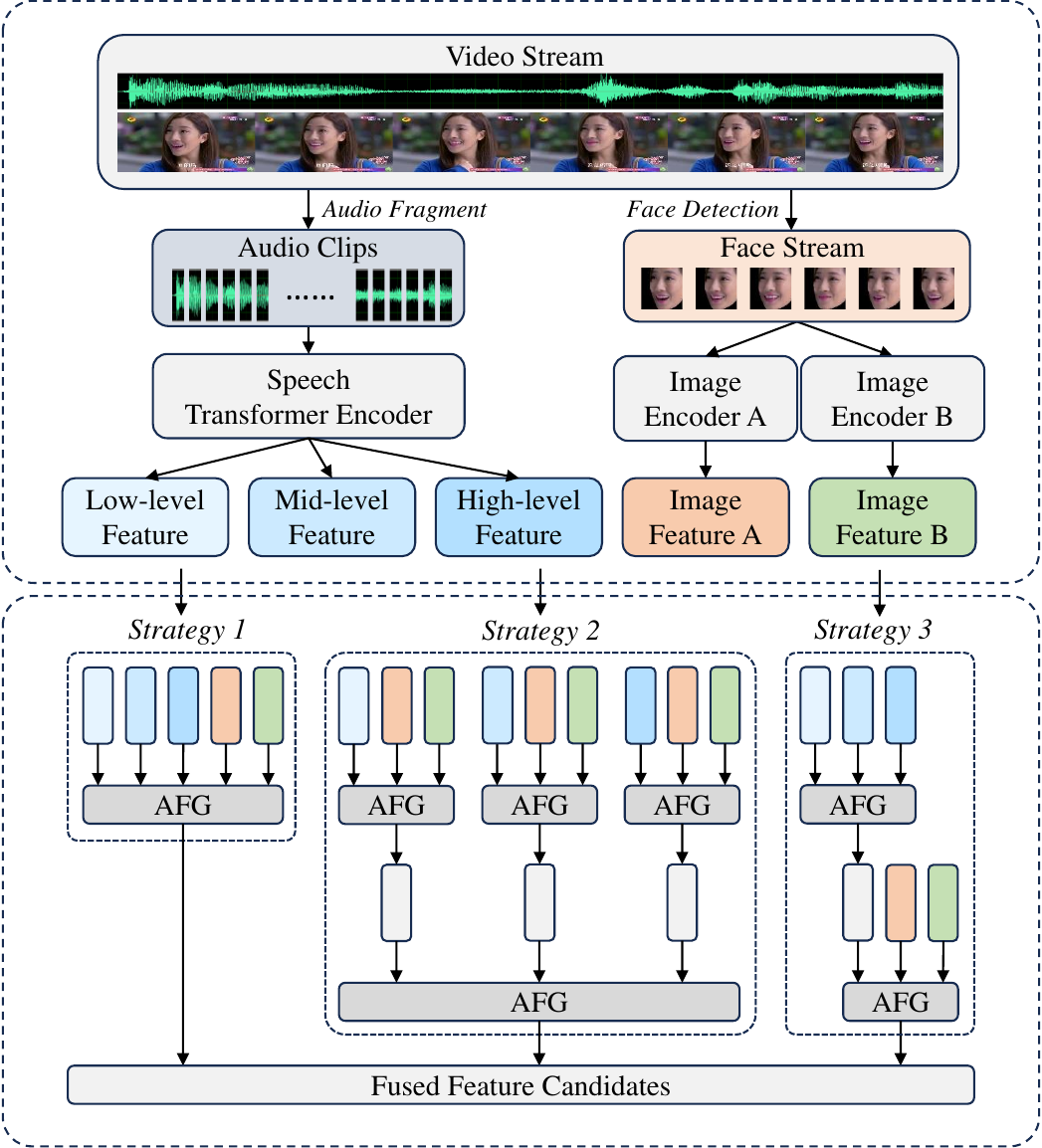}
  \caption{\centering{Deep acoustic and visual features extraction and the following three feature fusion structures. Structure of the attention-guided feature gathering (AFG) module is shown in the following figure.}}
  \label{fig:Feature extraction and fusion}
\end{figure}
\subsection{Features encoding and attention-guided fusion}
\begin{figure}[t]
  \centering
  \includegraphics[width=1.0\linewidth]{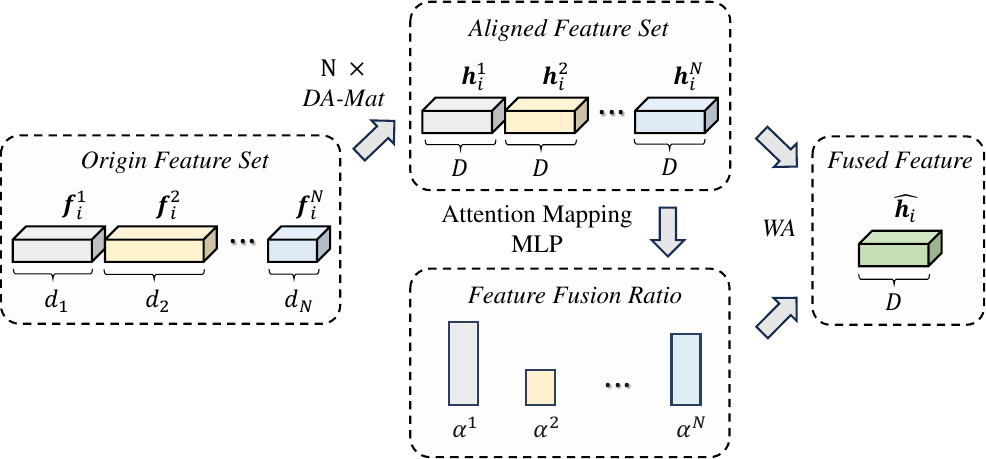}  \caption{\centering{Attention-guided feature gathering (AFG) module. DA-Mat is short for Dimension Align Matrix. WA is short for Weight Average.}}
  \label{fig:AFG}
\end{figure}
In our proposed architecture, we first extract deep features from pre-trained models as robust acoustic and visual representations of raw video segments. The details are illustrated in Figure~\ref{fig:Feature extraction and fusion}. Previous research has indicated that different layers in pre-trained speech model HUBURT~\cite{hubert} capture audio hidden states with distinctive characteristics~\cite{IASR}. The hidden states captured by layers closer to the front exhibit increased sensitivity to the acoustic features of the original audio, such as tone or frequency. In contrast, the hidden states captured by layers towards the back demonstrate a heightened sensitivity to the semantic information embedded within the audio. These different features with distinct acoustic information from the same audio segment can be complementary. Hence, for the acoustic modality, our framework incorporates different HUBURT layers to extract low-level, mid-level, and high-level audio features, forming a unified representation of the original audio. For visual modality, we first crop and align faces of raw video in each frame using the OpenFace~\cite{openface} toolkit. Then, we utilize various pre-trained models (such as MANet~\cite{MAnet} and ResNet-50~\cite{resnet}) to extract frame-level features and apply average encoding to compress them into video-level embeddings.  

Then, these different acoustic and visual representations will be fused together with deep feature fusion frameworks based on attention-guided feature gathering (AFG)~\cite{AFG}. In addition, features of text modality, however, are proved to be underperforming comparing with acoustic and visual features in this task~\cite{MER2023}
 so we do not take features from text encoders into consideration in our framework. The architecture of AFG  module is shown in Figure~\ref{fig:AFG}, whose principle is as follows:
\begin{equation}
    \boldsymbol{h}_i = {\rm Concat}(\boldsymbol{h}_i^1, \boldsymbol{h}_i^2,...,\boldsymbol{h}_i^N)
\end{equation}
\begin{equation}
    \boldsymbol{\alpha}_i={\rm Softmax}( \boldsymbol{h}_i^{T} \boldsymbol{W_\alpha} +  \boldsymbol{b_\alpha})
\end{equation}
\begin{equation}
   \hat{\boldsymbol{h}_i} =  \boldsymbol{h}_i  \boldsymbol{\alpha}_i
\end{equation}
where $ \boldsymbol{h}_i^1,\boldsymbol{h}_i^2,... ,\boldsymbol{h}_i^N \in \mathbb{R}^{D}$ are the aligned hidden states of sample $i$. $\boldsymbol{h}_i^{T}$ represents the transposed vector of $\boldsymbol{h}_i$, $ \boldsymbol{\alpha}_i \in \mathbb{R}^{N}$ is the attention score that indices the importance of different features.$ \boldsymbol{W_\alpha} \in \mathbb{R}^{D\times N}$ and $ \boldsymbol{b_\alpha} \in \mathbb{R}^ {N}$ are trainable attention matrix and bias. $\hat{\boldsymbol{h}_i} \in \mathbb{R}^{D}$ is the fused hidden state.

Three different feature fusion frameworks have been designed based on AFG~\cite{AFG}, as depicted in Figure~\ref{fig:Feature extraction and fusion}. In the first framework, we use AFG to fuse all features from acoustic and visual modalities in parallel. In the second framework, we fuse each acoustic representation with visual features to generate different audio-visual representations. These representations are then fused together, as they exhibit strong complementarity. While in the last framework, intra-modal fusion is firstly performed on the acoustic hidden states extracted by different layers of HUBURT~\cite{hubert} to create a unified acoustic representation, then inter-modal fusion is conducted based on the unified acoustic representation and visual representations.

\subsection{Joint decoding of discrete and dimensional emotions}

\begin{figure}[t]
  \centering
  \includegraphics[width=1.0\linewidth]{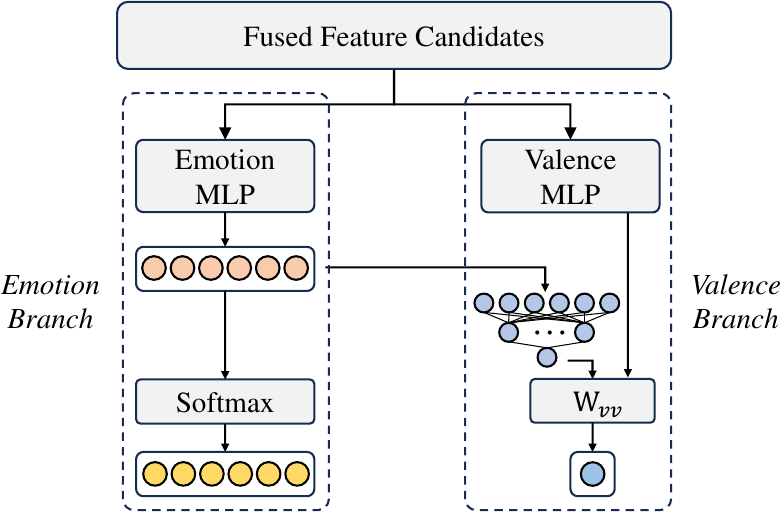}
  \caption{\centering{Joint decoding module of discrete and dimensional emotions. Fused feature candidates are the outputs of feature fusion module shown in Figure 1.}}
  \label{fig:JDEV}
\end{figure}

In our studies, we found there exists a relatively stable distribution between discrete and dimensional emotions \cite{multilabel1, multilabel2}. In other words, the discrete emotions can determine the dimensional emotions according to the following formula, where $e_i$ represents discrete emotions while $v_i$ represents dimensional emotions. $k$ represents categories of emotions.
\begin{equation}
    P(v_i) = \sum_{k=1}^{M} p(v_i|e_i=k)p(e_i=k)
\end{equation}
Therefore, our research incorporates a branch for discrete emotions judgment of dimensional emotions into the network structure. This branch establishes a mapping from the discrete space to the dimensional space, as depicted in Figure~\ref{fig:JDEV}.
The prediction of discrete and dimensional 
emotions by our multi-task framework is as follows: 
\begin{equation}
    \hat{\boldsymbol{e}_i}={\rm Softmax}(\tilde{\boldsymbol{e}_i})={\rm Softmax}(\hat{\boldsymbol{h}_i} \boldsymbol{W_e}+ \boldsymbol{b_e})
\end{equation}
\begin{equation}
    \hat{\boldsymbol{v}_i}={\rm Concat}( \tilde{\boldsymbol{v}_i}, \tilde {\boldsymbol{v}_i}^e) \boldsymbol{W_{vv}}+ \boldsymbol{b_{vv}}
\end{equation}
where $ \hat{\boldsymbol{h}_i}\in{\mathbb{R}^D}$ is the fused feature, $ \hat{\boldsymbol{e}_i}\in{\mathbb{R}^C}$ and $\hat{\boldsymbol{v}_i}\in{\mathbb{R}^1}$ are the estimated emotion and valence possibilities, respectively. $ \boldsymbol{W_{e}}\in{\mathbb{R}^{D\times C}}$, $ \boldsymbol{b_{e}}\in{\mathbb{R}^{D\times C}}$, $ \boldsymbol{W_{vv}}\in{\mathbb{R}^{2\times 1}}$ and $ \boldsymbol{b_{vv}}\in{\mathbb{R}^{2\times 1}}$ are trainable parameters. $ \tilde{\boldsymbol{v}_i}\in{\mathbb{R}}$ and $ \tilde {\boldsymbol{v}_i}^e\in{\mathbb{R}}$ are the estimated valence possibilities according to the fused state $\hat{\boldsymbol{h}_i}$ and emotion hidden state $\hat{\boldsymbol{e}_i}$ with trainable parameters $ \boldsymbol{W_{v}}\in{\mathbb{R}^{D\times 1}}$, $ \boldsymbol{b_{v}}\in{\mathbb{R}^{D\times 1}}$, $ \boldsymbol{W_{ev}}\in{\mathbb{R}^{C\times 1}}$ and $ \boldsymbol{b_{ev}}\in{\mathbb{R}^{C\times 1}}$ , as follows:
\begin{equation}
    \tilde{\boldsymbol{v}_i} = \hat{\boldsymbol{h}_i} \boldsymbol{W_v}+\boldsymbol{b_v}
\end{equation}
\begin{equation}
    \tilde{\boldsymbol{v}_i}^e = {\rm Tanh}(\tilde{\boldsymbol{e}_i}\boldsymbol{W_{ev}}+\boldsymbol{b_{ev}}) 
\end{equation}
During training, we use the cross-entropy (CE) loss as the classification loss, denoted as $\mathcal{L}_e$, for emotion prediction. The mean squared error (MSE) loss is adopted for valence prediction, denoted as $\mathcal{L}_v$. Moreover, we introduce dynamic weights to combine the two losses for better performance in the multi-task learning process. Inspired by the Uncertainty loss~\cite{uncertainty}, we introduce uncertainty loss weighting to $\mathcal{L}_e$ and $\mathcal{L}_v$, whose principle is as follows:
\begin{equation}
    \mathcal{L}_{ev} = \frac{1}{{\delta_1} ^2} \mathcal{L}_{e} + \frac{1}{2{\delta_2} ^2} \mathcal{L}_{v} + {\rm log}(1+\delta_1) + {\rm log}(1+\delta_2)
\end{equation}
where $\delta_1$ and $\delta_2$ are trainable uncertainty weights. We improved the regular loss term to ${\rm log}(1+\delta_1)$ and ${\rm log}(1+\delta_2)$ to avoid effects caused by enormous negative weights.

\section{Results and discussion}
We have conducted several experiments to evaluate the effectiveness of the proposed multimodal framework.

\subsection{Dataset and metric}
In this research, we conduct experiments on MER 2023 dataset~\cite{MER2023}. The dataset consists of 3373 labeled single-speaker video segments used as the training dataset. There are 411 and 412 unlabeled video segments for the test set in tracks 1 and 2, respectively. Same with the baseline~\cite{MER2023}, the combined metric of emotion classification and valence regression is chosen to evaluate the overall performance of discrete and dimensional emotions. 

\subsection{Implementations}
We have employed several data augmentation techniques specifically designed for emotional data on the training dataset for MER-NOISE sub-challenge. For the audio modality, we add speaker-independent noise with 7 different signal-to-noise ratios (from 5dB to 11dB with a step of 1dB) from the $speech$ subset of MUSAN~\cite{musan} to simulate audio segments of various qualities. For the visual modality, we perform various image transformations to augment the data, including variations in brightness (e.g., solarize), changes in articulation (e.g., blur), alterations in position (e.g., rotate), and modifications in image content (e.g., cutout).

We also conduct decision-level fusion on predictions of multimodal emotion recognition systems with three fusion strategies. The fused predictions are as follows:
\begin{equation}
\hat{\boldsymbol{e}}=k_1\hat{\boldsymbol{e}_1}+k_2\hat{\boldsymbol{e}_2}+k_2\hat{\boldsymbol{e}_3}
\end{equation}
\begin{equation}        
\hat{\boldsymbol{v}}=k_1\hat{\boldsymbol{v}_1}+k_2\hat{\boldsymbol{v}_2}+k_3\hat{\boldsymbol{v}_3}
\end{equation}
where $\hat{\boldsymbol{e}_i}$ and $\hat{\boldsymbol{v}_i}$ represents the emotion and valence prediction vectors from three different multimodal emotion recognition systems. $k_1,k_2,k_3 (k_1+k_2+k_3=1)$ represents the weighted factors of three systems respectively. The final decisions of emotion and valence are based on the posterior probability outputs of separate systems with different feature fusion encoders.

\begin{figure}[t]
    \centering
    \includegraphics[width=\linewidth]{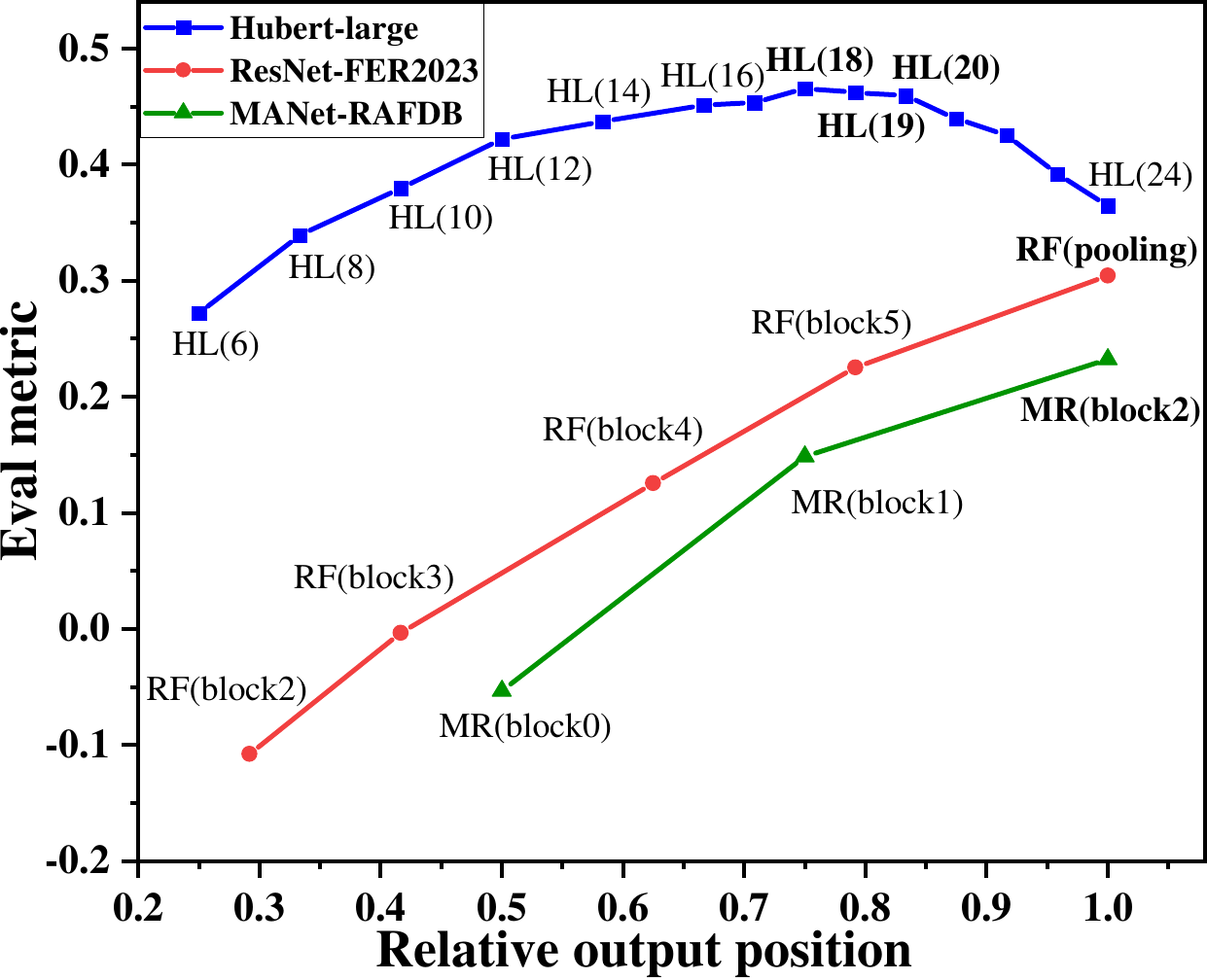}
    \caption{\centering{Performance comparison of unimodal systems using different deep features as input.}}
    \label{fig:unimodal}
\end{figure}

\subsection{Performance comparison of unimodal systems}
We compared the performance of different unimodal emotion recognition systems. Different unimodal systems utilizing different layers of the same pre-trained model as deep feature encoders have also been considered.

\begin{table*}[t]
\caption{\centering{Performance comparison of multimodal systems with different deep features. "Dis", "Dim", and "Com" denote the performance on discrete, dimensional, and combined metrics. JDEV is short for Joint Decoding for Emotion and Valence.}}
\label{tab:multimodal}
\resizebox{2.1\columnwidth}{!}{\begin{tabular}{@{}cc|cc|cc|cccccccc@{}}
\toprule
\multicolumn{2}{c|}{Feature Encoder} & \multicolumn{2}{c|}{MR+RF} & \multicolumn{2}{c|}{HL(18)+HL(19)+HL(20)} & \multicolumn{8}{c}{\textbf{HL(18)+HL(19)+HL(20)+MR+RF}} \\ \midrule
\multicolumn{2}{c|}{Fusion Strategy} & \multicolumn{2}{c|}{1} & \multicolumn{2}{c|}{1} & \multicolumn{2}{c|}{1} & \multicolumn{2}{c|}{2} & \multicolumn{2}{c|}{3} & \multicolumn{2}{c}{\textbf{1+2+3 fused}} \\ \midrule
\multicolumn{2}{c|}{Decoder} & \multicolumn{1}{c|}{Baseline} & JDEV & \multicolumn{1}{c|}{Baseline} & JDEV & \multicolumn{1}{c|}{Baseline} & \multicolumn{1}{c|}{JDEV} & \multicolumn{1}{c|}{baseline} & \multicolumn{1}{c|}{JDEV} & \multicolumn{1}{c|}{Baseline} & \multicolumn{1}{c|}{JDEV} & \multicolumn{1}{c|}{Baseline} & \textbf{JDEV} \\ \midrule
\multirow{3}{*}{Train$\&$Val} & Dis($\uparrow$) & \multicolumn{1}{c|}{0.6085} & 0.6170 & \multicolumn{1}{c|}{0.6995} & 0.7051 & \multicolumn{1}{c|}{0.7743} & \multicolumn{1}{c|}{0.7811} & \multicolumn{1}{c|}{0.7769} & \multicolumn{1}{c|}{0.7789} & \multicolumn{1}{c|}{0.7735} & \multicolumn{1}{c|}{0.7795} & \multicolumn{1}{c|}{0.7865} & \textbf{0.7936} \\
 & Dim($\downarrow$) & \multicolumn{1}{c|}{1.2291} & 1.1890 & \multicolumn{1}{c|}{0.9568} & 0.9297 & \multicolumn{1}{c|}{0.6750} & \multicolumn{1}{c|}{0.6176} & \multicolumn{1}{c|}{0.6714} & \multicolumn{1}{c|}{0.6339} & \multicolumn{1}{c|}{0.6659} & \multicolumn{1}{c|}{0.6315} & \multicolumn{1}{c|}{0.6364} & \textbf{0.6138} \\
 & Com($\uparrow$) & \multicolumn{1}{c|}{0.3012} & 0.3198 & \multicolumn{1}{c|}{0.4603} & 0.4727 & \multicolumn{1}{c|}{0.6056} & \multicolumn{1}{c|}{0.6267} & \multicolumn{1}{c|}{0.6091} & \multicolumn{1}{c|}{0.6204} & \multicolumn{1}{c|}{0.6070} & \multicolumn{1}{c|}{0.6216} & \multicolumn{1}{c|}{0.6247} & \textbf{0.6402} \\ \midrule
MER-MULTI & Com($\uparrow$) & \multicolumn{1}{c|}{0.3072} & 0.3171 & \multicolumn{1}{c|}{0.4997} & 0.5184 & \multicolumn{1}{c|}{0.6675} & \multicolumn{1}{c|}{0.6779} & \multicolumn{1}{c|}{0.6550} & \multicolumn{1}{c|}{0.6656} & \multicolumn{1}{c|}{0.6618} & \multicolumn{1}{c|}{0.6693} & \multicolumn{1}{c|}{0.6787} & \textbf{0.6846} \\ \midrule
MER-NOISE & Com($\uparrow$) & \multicolumn{1}{c|}{0.2984} & 0.3074 & \multicolumn{1}{c|}{0.4933} & 0.4934 & \multicolumn{1}{c|}{0.6110} & \multicolumn{1}{c|}{0.6178} & \multicolumn{1}{c|}{0.6066} & \multicolumn{1}{c|}{0.6166} & \multicolumn{1}{c|}{0.6058} & \multicolumn{1}{c|}{0.6125} & \multicolumn{1}{c|}{0.6162} & \textbf{0.6303} \\ \bottomrule
\end{tabular}}
\end{table*}
Figure~\ref{fig:unimodal} presents the performance diversity of different layers of audio and visual systems on Train$\&$Val. HL($i$) indicates using the output of $i$-th layer of HUBURT-large~\cite{hubert} model as acoustic feature. RF($j$) and MR($j$) indicate using the output of $j$-th block of ResNet-FER2013~\cite{resnet} or MANet-RAFDB~\cite{MAnet} as encoder output. Relative output position indicates the relative positions that the output layer in the whole pre-trained model (HUBURT-large~\cite{hubert}, ResNet~\cite{resnet} or MANet~\cite{MAnet}).

Among various acoustic features, HL(18), HL(19), HL(20) outperform others, confirming that the mid-level features from HUBERT-large~\cite{hubert} model is more suitable for emotion recognition. For visual modality, high-level features generated by the last block of ResNet~\cite{resnet} and MANet~\cite{MAnet} outperform others. In general, acoustic features performs better than visual features in this task.  

\subsection{Performance comparison of multimodal systems}
In this section, we perform multimodal fusion based on 3 well-performing acoustic features HL(18), HL(19), HL(20) and 2 visual features RF(pooling) and MR(block2). We conduct both intra-modal and inter-modal fusion. Three fusion strategies are conducted separately in inter-modal fusion. Then, the baseline\cite{MER2023} and the proposed joint decoding manner are utilized separately to obtain predictions for emotion and valence. The results are shown in Table~\ref{tab:multimodal}. 

The results indicate that features from different layers of the same model can be complementary, as the fusion metric improves by 1-2 percent points comparing with single-layer feature. Furthermore, the results demonstrate that incorporating features from different modalities significantly improves performance. The final multimodal system achieves a score of 0.6846 tested on MER-MULTI, which is a 16.6 percent improvement to the unimodal system.

Interestingly, we observed a significant gain on MSE loss for dimensional emotion regression when utilizing JDEV. This phenomenon suggests that the relatively reliable results of emotion classification can assist in improving the accuracy of dimensional valence regression by joint decoding.

Among the three proposed fusion strategies, the parallel fusion strategy 1 achieves the highest metric of 0.6779 on MER-MULTI. Additionally, when combining all three strategies on posterior probability level, we obtain an additional score gain of 0.67 percent points when tested on MER-MULTI sub-challenge. 

\section{Conclusions}
In this paper, we propose a hierarchical audio-visual information fusion framework for recognizing both discrete and dimensional emotions. Three different feature fusion encoders are designed for deep feature fusion. In the decoding stage, we introduce a joint decoding structure for emotion classification and valence regression. In addition, a multi-task loss is also designed as optimizer for the whole process. Finally, by combining three different structures on posterior probability level, we obtain the final predictions of both emotion and valence. When tested on the dataset of MER 2023, our final system ranks third on the MER-MULTI sub-challenge.

\section{Acknowledgements}
This work was supported by the National Natural Science Foundation of China under Grants No. 62171427.

\bibliographystyle{ACM-Reference-Format}
\balance
\bibliography{MER_camera_ready}

\end{document}